\documentclass{article}
\usepackage{spconf,amsmath,graphicx}

\usepackage{amssymb}
\usepackage{booktabs}
\usepackage{combelow}
\usepackage{microtype}
\usepackage{multirow}
\usepackage{xspace}
\usepackage{url}

\usepackage{mymacros}
\usepackage{tikz}

\usetikzlibrary{backgrounds}
\usetikzlibrary{decorations.pathreplacing}
\usetikzlibrary{calc}

\def\pb{\mathbf{p}}
\def\tb{\mathbf{t}}
\def\thetab{\mathbf{\theta}}
\def\st{s^{(\mathsf{t})}}
\def\sw{s^{(\mathsf{w})}}

\def\softmax{\mathrm{softmax}}
\title{
    An evaluation of word-level confidence estimation \\for end-to-end automatic speech recognition
}
\name{
    Dan Onea\cb{t}\u{a}$^1$,
    Alexandru Caranica$^1$,
    Adriana Stan$^2$,
    Horia Cucu$^1$
}
\address{
    $^1$University \textsc{Politehnica} of Bucharest, Romania \\
    $^2$Technical University of Cluj-Napoca, Romania
}

\begin{document}
%\ninept
%
\maketitle
\begin{abstract}
    Quantifying the confidence (or conversely the uncertainty) of a prediction is a highly desirable trait of an automatic system,
    as it improves the robustness and usefulness in downstream tasks.
    In this paper we investigate confidence estimation for end-to-end automatic speech recognition (ASR).
    Previous work has addressed confidence measures for lattice-based ASR,
    while current machine learning research mostly focuses on confidence measures for unstructured deep learning.
    However, as the ASR systems are increasingly being built upon deep end-to-end methods,
    there is little work that tries to develop confidence measures in this context.
    We fill this gap by providing an extensive benchmark of popular confidence methods on four well-known speech datasets.
    There are two challenges we overcome in adapting existing methods:
    working on structured data (sequences) and
    obtaining confidences at a coarser level than the predictions (words instead of tokens).
    Our results suggest that a strong baseline can be obtained by
    scaling the logits by a learnt temperature,
    followed by estimating the confidence as the negative entropy of the predictive distribution and,
    finally, sum pooling to aggregate at word level.
\end{abstract}
\section{Introduction}
\label{sec:intro}

Reasoning under uncertainty is one of the tenets of intelligence.
The first step towards this goal is to endow systems with reliable uncertainty estimates of their predictions.
% We want to endow our automatic systems with capability.
% In this work we consider uncertainty estimation for the task of automatic speech recognition (ASR).
Ideally, the larger the uncertainty the more likely the prediction is erroneous.
Alternatively, one can solve the complementary problem of confidence estimation---%
in this case, the more confident a prediction, the more likely the output is correct.
% The two problems are complementary, as the probability of the output of being correct is the complement of it being incorrect.

In the context of automatic speech recognition (ASR)
confidence estimation can be of crucial importance for many end-user applications,
as it improves the robustness of the systems in safety-critical tasks,
helps avoiding errors in human-computer dialogue systems and
facilitates manual corrections in audio transcription tasks by flagging the errors.
%Uncertainty estimates not only improve the robustness of the systems in safety-critical tasks,
%but they can also help performance in a number of scenarios.
% helps performance in a number of more complex scenarios.
Moreover, previous research has leveraged confidence estimates for a number of downstream tasks:
propagating uncertainties for automatic speech translation \cite{sperber2017emnlp},
selecting confident predictions for self-training \cite{vesely2013asru},
manually annotating the less confident predictions for active learning \cite{yu2010csl}.
% For example the uncertainty over the ASR transcript can improve on downstream tasks,
% such as, automatic speech translation

In this paper we consider confidence estimation for \textit{end-to-end} ASR systems,
also known as lattice-free speech recognition \cite{hadian2018interspeech}.
End-to-end models for ASR are gaining traction recently as
their performance matches the one of classical ASR and
have the additional benefits of being conceptually simple and allowing unified training \cite{luscher2019interspeech,tuske2019interspeech,karita2019asru}.
However, there is surprisingly little work on confidence estimation for end-to-end speech recognition systems,
most of the ongoing research on confidence estimation being carried on computer vision tasks (image classification or segmentation).
% Challenges
We believe that there are two main challenges to developing confidence scoring methods for ASR systems:
the structured output and the granular predictions (\eg, tokens or graphemes versus words).
% the structured output of the ASR systems and
% the more granular predictions than what one is usually interested in (\eg, tokens versus words).

ASR systems are structured models (mapping sequences to sequences) as opposed to usual recognition networks (such as, image classification) whose output is a single label.
The sequential nature of the output imposes a decoding step,
which complicates not only the prediction but also the confidence scoring algorithm,
as we need estimate the confidence in an auto-regressive context (the already predicted sequence).
For this reason, we fix the predictions based on a pre-trained ASR and
apply the confidence scoring methods on top of token probabilities, which are conditioned on the fixed transcript.
% The length of the prediction is variable.

In order to enable open vocabulary predictions,
end-to-end ASR systems usually use subword tokens to represent the output (byte-pair encoded tokens or even graphemes). % TODO Reference
However, given that the tokens lack semantics,
for many downstream applications we are interested in estimating the confidence of words.
To this end, we explore ways of aggregating the token-level uncertainty measures to the larger units, corresponding to words;
in fact, the presented techniques can be extended to even coarser predictions, such as sentence or utterance level.

% - By learning the aggregation function we not only improve the ranking of the words,
% but we also obtain a valid probability distribution.
% To show that this is better calibrated than performing average or product of initial probabilities.
% - The same methodology can be applied to assign a confidence to the entire sentence.

In this context, our main contributions are the following:
\ia we adapt several state-of-the-art uncertainty estimation methods to the end-to-end ASR pipeline;
\ib we propose and evaluate aggregation techniques to obtain user-relevant confidence estimates (\ie word-level);
\ic we perform a thorough evaluation on multiple speech benchmark datasets.
To the best of our knowledge, this is the first study that provides an in-depth analysis of confidence measures for end-to-end ASR.
%Contributions:
%\ia adapt uncertainty methods for this task and evaluate aggregation techniques to obtain word-level confidence estimates.
% - Maximum probability, see the baseline paper of Hendrycks and Gimpel (2017).
% - Entropy, used in the dropout paper of Gal and Ghahramani (2015).
%\ib provide a thorough evaluation of the methods on multiple benchmark datasets.
%\ic strive for reproducible research.

\section{Related work}
\label{sec:related-work}

In this section we review two lines of research that are related to our work.
% They roughly fall in two lines of research:
% confidence scoring for speech recognition systems and
% confidence scoring for deep learning.

\textit{Confidence scoring for speech recognition.}
Most prior work on confidence scoring for ASR targets classical systems based on the HMM-GMM paradigm.
These methods first extract a set of features from the decoding lattice, acoustic or language model,
and then train a classifier to predict whether the transcription is correct or not.
Typical examples of features include
log-likelihood of the acoustic realization, language model score, word duration, number of alternatives in the confusion network \cite{kemp1997eurospeech,weintraub1997icassp,hazen2002csl}.
More recently, Swarup \etal have augmented the feature set with deep embeddings of the input audio and the predicted text \cite{swarup2019interspeech},
% based on deep recurrent architecture,
while Errattahi \etal have shown that the benefits of domain adaptation on the extracted features \cite{errattahi2018slt}.
% the language model features can be refined by adapting the language model to the output transcript.
The classifiers employed by the confidence scoring methods range from
conditional random fields \cite{seigel2013phd,cortina2016csl} and
multiple layer perceptrons \cite{kalgaonkar2015icassp} to
bidirectional recurrent neural networks \cite{ogawa2017speechcomm,delagua2018,li2019icassp}.

% \cite{swarup2019interspeech}
% these features are integrated with more standard decoder features in the confidence scoring model,
% which is implemented as a logistic regression classifier.
%
% % TODO Hain's article on error detection from SLT
% \cite{errattahi2018slt}
% also augment classic decoder features (from the confusion network) with information from the neural language model.
% Here the features include probability according to the language model and forward and backward distance.
% This probabilities are further refined by running an adaptation step based on the output transcript.

\textit{Confidence scoring in end-to-end systems.}
The baseline method for confidence estimation in neural networks is to use directly the probability of the most-likely prediction \cite{hendrycks2016iclr}.
However the neural networks tend to be overconfident and the probability estimates can be improved through temperature scaling \cite{hinton2015arxiv},
which typically leads to better calibration \cite{guo2017icml,ashukha2020iclr}.
The most promising direction in terms of simplicity and usefulness involves Monte Carlo estimation:
Gal and Ghahramani use dropout at test time to obtain multiple predictions, which are then averaged \cite{gal2016icml},
while Lakshminarayanan \etal average the predictions over an ensemble of networks usually trained with different initializations \cite{lakshminarayanan2017nips}.
The latter has been show to be very reliable on challenging out-of-domain datasets \cite{ovadia2019nips},
but coming at a high cost \cite{ashukha2020iclr}.
The literature on general confidence scoring is rich and continually evolving;
the most interesting research avenues involve
Bayesian averaging \cite{maddox2019nips},
generative models \cite{nalisnick2018iclr,che2019arxiv},
input perturbations \cite{liang2018iclr,thulasidasan2019nips},
exploiting inner activations \cite{corbiere2019nips,chen2019aistats}.

% [a] Maddox, Wesley J., et al. &quot;A simple baseline for Bayesian uncertainty in deep learning.&quot; NeurIPS, 2019.
% [b] Nalisnick, Eric, et al. &quot;Do deep generative models know what they don&#039;t know?.&quot; ICLR, 2019.
% [c] Liang, Shiyu, et al. &quot;Enhancing the reliability of out-of-distribution image detection in neural networks.&quot; ICLR, 2018.
% [d] Thulasidasan, Sunil, et al. &quot;On mixup training: Improved calibration and predictive uncertainty for deep neural networks.&quot; NeurIPS, 2019.
% A different approach to confidence scoring is to learn a classifier (typically another neural network)

At the intersection of the two lines of research,
there is the recent work of Malinin and Gales \cite{malinin2020arxiv},
which similar to us addresses the task of confidence estimation for end-to-end ASR systems.
However, they are concerned with token and sentence uncertainty estimation,
while we are interested in estimation at word level,
and, consequently, provide more focus on the aggregation techniques.
Furthermore, they employ ensembles as their primary method of confidence estimation,
while we also evaluate temperature scaling and dropout methods.
Dropout was previously used for obtaining confidence scores for ASR \cite{vyas2019icassp}, but our approaches differ:
% the method is different from our approach. 
in \cite{vyas2019icassp} multiple hypotheses are generated via dropout and then word confidences are assigned based on their frequency of appearance in the aligned hypotheses;
in contrast, we aggregate the posterior probabilities and not the hypotheses, which simplifies the procedure as it avoids the alignment step.

\section{Methodology}
\label{sec:methodology}

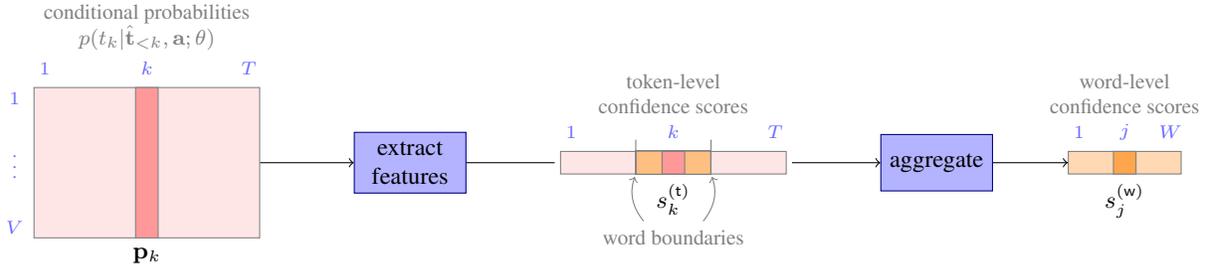
\begin{figure*}
    \centering
    \begin{tikzpicture}[%
            font=\small,
            auto,
            block/.style={text width=1.25cm, minimum height=0.75cm, minimum width=0.5cm, align=center},
            mylabel/.style={align=center, color=gray, font=\footnotesize, label distance=3.75mm},
            m dim/.style={color=blue!60, font=\scriptsize},
        ]
        \def\posa{0}
        \def\posb{3.5}
        \def\posc{7.0}
        \def\posd{10.5}
        \def\pose{13}
        \node[minimum height=2cm, minimum width=3cm, fill=red!10, draw=gray, label={[mylabel]conditional probabilities\\$p(t_k|\hat\tb_{<k}, \ab; \thetab)$}] at (\posa, 0) (probas) {};
        \node[minimum height=2cm, minimum width=0.3cm, fill=red!40, draw=gray, label={below:$\pb_k$}] at (\posa, 0) (proba) {};

        \node[block, fill=blue!30, draw=blue!50!black] at (\posb, 0) (feat) {extract features};

        \node[minimum height=0.3cm, minimum width=3cm, fill=red!10, draw=gray, label={[mylabel]token-level\\confidence scores}] at (\posc, 0) (scores token) {};
        \begin{scope}
            \clip[postaction={fill=orange!50, draw=gray, line width=0.22mm}] ($(scores token.south) + (-0.50, 0)$) rectangle ($(scores token.north) + (+0.50, 0)$);
        \end{scope}
        \node[minimum height=0.3cm, minimum width=0.3cm, fill=red!40, draw=gray, label={below:$\st_k$}] at (\posc, 0) (score token) {};

        \node[block, fill=blue!30, draw=blue!50!black] at (\posd, 0) (agg) {aggregate};

        \node[minimum height=0.3cm, minimum width=1.5cm, fill=orange!30, draw=gray, label={[mylabel]word-level\\confidence scores}] at (\pose, 0) (scores word) {};
        \node[minimum height=0.3cm, minimum width=0.3cm, fill=orange!70, draw=gray, label={below:$\sw_j$}] at (\pose, 0) (score word) {};

        \node[m dim] at ($(probas.north west) + (-0.25, -0.15)$) {$1$};
        \node[m dim] at ($(probas.west)       + (-0.25,  0.05)$) {$\vdots$};
        \node[m dim] at ($(probas.south west) + (-0.25,  0.15)$) {$V$};

        \node[m dim] at ($(probas.north west) + ( 0.15,  0.25)$) {$1$};
        \node[m dim] at ($(probas.north)      + ( 0.00,  0.25)$) {$k$};
        \node[m dim] at ($(probas.north east) + (-0.15,  0.25)$) {$T$};

        \node[m dim] at ($(scores token.north west) + ( 0.15,  0.25)$) {$1$};
        \node[m dim] at ($(scores token.north)      + ( 0.00,  0.25)$) {$k$};
        \node[m dim] at ($(scores token.north east) + (-0.15,  0.25)$) {$T$};

        \node[m dim] at ($(scores word.north west) + ( 0.15,  0.25)$) {$1$};
        \node[m dim] at ($(scores word.north)      + ( 0.00,  0.25)$) {$j$};
        \node[m dim] at ($(scores word.north east) + (-0.15,  0.25)$) {$W$};

        \draw[gray] ($(scores token.south) + (-0.50, 0)$) -- ($(scores token.north) + (-0.50, 0.15)$);
        \draw[gray] ($(scores token.south) + (+0.50, 0)$) -- ($(scores token.north) + (+0.50, 0.15)$);

        \node[below of=score token,font=\footnotesize, color=gray] (wb label) {word boundaries};

        % \node[m dim] at (\posa, -1.00) {$1 \times T$};
        % \node[m dim] at (\posc, -1.50) {$D \times W$};
        % \node[m dim] at (\pose, -1.25) {$K \times F$};
        % \node[m dim] at (14,    -1.25) {$K \times 1$};

        \draw[->] (probas) -- (feat);
        \draw[shorten >=2pt] (feat) -- (scores token);
        \draw[shorten <=2pt, ->] (scores token) -- (agg);
        \draw[->] (agg) -- (scores word);
        \draw[->, gray, shorten >=1pt] (wb label) to [bend left]  ($(scores token.south) + (-0.50, 0)$);
        \draw[->, gray, shorten >=1pt] (wb label) to [bend right] ($(scores token.south) + (+0.50, 0)$);
    \end{tikzpicture}
    \caption{%
        Overview of the confidence scoring procedure.
        From an end-to-end ASR system we obtain probabilities $\pb_k$
        of the $k$-th token given an utterance $\ab$ and previously predicted tokens $\hat\tb_{<k}$.
        Based on these probabilities we extract token-level confidence scores $\st$,
        which we then aggregate to obtain scores at word level $\sw$.
        % (the orange region from the middle picture shows tokens belonging to the same word).
        The size of the token vocabulary is denoted by $V$,
        the number of tokens is denoted by $T$ and
        the number of words by $W$.
    }
    \label{fig:pipeline}
\end{figure*}

This section presents the confidence estimation methodology and proposed ways of improving them.
We first start with a description of the setup and the involved notation.

We consider a sequence-to-sequence model that maps an audio sequence $\ab$ to a sequence of tokens $\tb = (t_1,\cdots,t_T)$.
The model is specified by the parameters $\thetab$, which are learned by minimizing losses such as the CTC or KL divergence on the training set.
At test time the model outputs probabilities for the next token $k$ in an auto-regressive manner $p(t_k|\hat\tb_{<k}, \ab; \thetab)$ based on the already predicted tokens $\hat\tb_{<k}$.
These probabilities are used for performing decoding via beam search to obtain the most likely sequence of tokens.
Given that the conditioned output probability is a distribution over the $V$ tokens in the vocabulary,
we denote it by a $V$-dimensional vector, $\pb_k$.

The main assumption of our methodology is the availability of a probability distribution over each token.
This criterion is satisfied by most end-to-end ASR architectures including the
RNN transducer \cite{graves2012arxiv},
recurrent neural aligner \cite{sak2017interspeech},
attention-based encoder-decoder \cite{bahdanau2016icassp} and
2D LSTMs \cite{bahar2019icassp}.

\subsection{Confidence estimation}
\label{subsec:confidence-estimation}

Our goal is to obtain a confidence score for each word in the output transcript of the ASR.
We achieve this in two steps.
First, using the posterior probabilities at each time step $\pb_k$,
we extract features to encode the confidence score of each token $\st_k$.
Second, we aggregate the token-level scores into word-level confidence scores $\sw_j$,
based on the word boundaries.
Next we detail these two steps; see also Figure \ref{fig:pipeline}.
% We assume we are given only the scores (posterior probabilities) from the ASR.
% Instead, as a first step we strive approaches that are practical and do not rely on anything than existing models.
% These probabilities are conditioned on the previously predicted units and the units usually represent sub-word information (characters or tokens);
% we use the following notation for the scores $\pb_k = \log p(t_{k}|\mathbf{t}_{1:k-1}, \mathbf{a})$.

\textbf{Feature extraction.}
To measure the confidence in a prediction at token level we use two variants:
\begin{itemize}
    \item Log probability (log-proba) of the most probable prediction given by classifier, that is $\st = \log\max\pb$.
        This type of feature has been shown to yield a strong baseline for the related tasks of misclassification and out-of-distribution detection \cite{hendrycks2016iclr}.
    \item Negative entropy (neg-entropy) computed over the vocabulary of tokens at each time step, that is $\st = \pb^{\intercal}\log\pb$.
        A large entropy means a large uncertainty or, conversely, a large negative entropy implies a confident prediction.
\end{itemize}

\textbf{Aggregation.}
To obtain word-level features from the token-level ones,
we experiment with three types of aggregation functions: sum, average, minimum.
Since both proposed features are negative, summing across tokens will result in smaller values and, hence, in lower confidences;
this behaviour can be desirable as longer words are more likely to be erroneous (see Figure \ref{fig:error-vs-length}).
Also, when we sum the log probability of the tokens,
we obtain a word-level score corresponding to the log probability of the entire sequence.
% For example the sum of log probabilities is equivalent to obtaining the probability of the sequence.
Taking minimum is justified by the fact that we might want a low confidence if at least one of the tokens has low confidence.

\subsection{Improving the token probabilities}

We propose three ways to make the token probabilities reliable:
temperature scaling, dropout and ensembles of models.
Our assumption is that by improving the token probabilities,
we also improve the word-level scores.

\textbf{Temperature scaling} \cite{hinton2015arxiv, guo2017icml}
consists of dividing the logit activations (pre-softmax values) by a scalar $\tau$ (known as temperature).
The value of $\tau$ ranges from zero to infinity and it controls the shape of the distribution:
when $\tau \rightarrow 0$ we obtain a uniform distribution,
when $\tau \rightarrow \infty$ we obtain a Dirac distribution on the most likely output.
% Instead of extracting pre-defined features such as the maximum probability or the entropy,
% we parameterize the process and learn a factor $\tau$ that controls the temperature scaling --
Based on $\tau$ we update token-level probabilities $\pb$ at each time step $k$, as follows:
\begin{equation}
  \pb'_k = \softmax(\log(\pb_k) / \tau).
\end{equation}

We then extract features $\st$ on the updated probabilities $\pb'$,
aggregate them into the word-level score $\sw$ and,
finally, classify the word as either correct or incorrect:
\begin{equation}
    P(\mathrm{correct}) = \sigma(\alpha \cdot \sw + \beta).
\end{equation}
% where $\sigma$ is the sigmoid function and $\sw$ is a word-level score obtained from aggregating the token-level scores $\st$.

The variables $\alpha$, $\beta$ and $\tau$ are parameters and are learnt by optimizing the cross-entropy loss on a validation set.
The labels are set at word level by aligning at the groundtruth text with the transcription.
Note that the parameters $\alpha$ and $\beta$ are not changing the ranking of the predictions,
but allow us to learn a calibrated confidence model.

\textbf{Dropout} \cite{srivastava2014jmlr}
is a technique that masks out random parts of the activations in a network,
making the network less prone to overfitting.
% It has been shown it makes the network more robust and it is used for regularization.
In \cite{gal2016icml} it has been observed that the dropout induces a probability distribution over the weights of the network
and can be consequently used for approximate Bayesian inference.
We follow this idea and average the token probabilities obtained over multiple runs of dropout:
\begin{equation}
    \pb_k' = \frac{1}{N}\sum_{n} \hat{\pb}_k
\end{equation}
where $\hat{\pb}$ specifies the dropout prediction.
While the original work \cite{gal2016icml} employed entropy as a confidence measure,
there is no reason not use other uncertainty features;
we use the updated probabilities $\pb'$ to extract both log-proba and neg-entropy features.
% it is by no means restricted to this usage, and can be also applied on the original probabilities.
% The updated probabilities are then used to extract either of the

\textbf{Ensembles} \cite{lakshminarayanan2017nips}
are based on the same idea of averaging predictions from multiple sources,
but in this case the set of weights come from independently trained networks (different random seeds used in initialization and batch selection).%,
%which differ in their initialization.
In our case, we average the token predictions over the models:
\begin{equation}
    \pb_k' = \frac{1}{N}\sum_{n} p(t_k|\hat\tb_{<k}, \ab; \thetab_n),
\end{equation}
where $\left\{\thetab_n\right\}_{n=1}^N$ specifies the ensemble of models.
Note that we need to have the same context $\hat\tb_{<k}$ for all models in the ensemble,
so we use the one given by a pre-trained model.

The three presented approaches can be combined;
for example,
we can first update the probabilities using temperature scaling
then average them using dropout.
In the experimental section we will evaluate all these combinations.

\section{Experimental setup}
\label{sec:experimental-setup}

In this section, we describe the datasets used for evaluation, the ASR systems for which we build confidence estimates, and the evaluation metrics.

\subsection{Datasets}
\label{subsec:datasets}

We have opted for multiple publicly-available and widely-used datasets for our experimental setup.

\textit{LibriSpeech} \cite{panayotov2015icassp}
is a corpus of approximately 1000 hours of read audiobooks derived from the LibriVox project.
% prepared by Vassil Panayotov with the assistance of Daniel Povey.
% The data is derived from read audiobooks, from the LibriVox project, and has been carefully segmented and aligned.
% The data was derived from the LibriVox project and has been carefully segmented and aligned.
% The main ASR system was trained on LibriSpeech \cite{panayotov2015icassp}.
We use the dataset for both training the ASR and evaluating the confidence scoring:
for training we use the three splits \texttt{clean100}, \texttt{clean360} and \texttt{other500},
while for development and evaluation we use the standard \texttt{clean} and \texttt{other} splits.
% For training, we included the following sets, from the database:
% \textit{train\_clean100}, \textit{train\_clean360} and \textit{train\_other500}, in order to obtain \textit{train\_960} set,
% reaching the maximum number of training hours offered by the dataset.
% split Evaluation is done on standard splimultiple datasets, as follows.
% From LibriSpeech, we used \textit{dev\_clean} and \textit{dev\_other} subsets.

\textit{TED-LIUM 2} \cite{rousseau2014lrec} consists of talks and their transcripts collected from the TED website.
We use the dataset for evaluation and consequently employ only the predefined \texttt{dev} and \texttt{test} subsets.

\textit{CommonVoice} \cite{ardila2020lrec} is a collaborative dataset of short transcripts that are read by people across the world;
we use the first release of the dataset.%
\footnote{\url{https://common-voice-data-download.s3.amazonaws.com/cv_corpus_v1.tar.gz}}
The data is used for evaluation and we defined \texttt{dev} and \texttt{test} subsets by choosing 10\% random samples for each of them.

\begin{table}
    \center
    \caption{
        Size of the datasets (\texttt{test} split) used for confidence estimation evaluation.
    }
    \begin{tabular}{crr}
        \toprule
        dataset     & no. utts. & duration \\
        \midrule
        Libri clean & 2.6K & 5.4 h \\
        Libri other & 2.9K & 5.3 h \\
        TED         & 1.1K & 2.6 h \\
        CommonVoice &  66K & 72 h \\
        \bottomrule
    \end{tabular}
    \label{tab:data-size}
\end{table}

Table \ref{tab:data-size} presents the \texttt{test} size of each evaluation dataset.

% Table \ref{tab:evalsets} summarizes the decoding sets used, also declaring our naming convention used throughout the paper.
%
% \begin{table*}
%     \center
%     \small
%     \setlength{\tabcolsep}{3pt}
%     \caption{%
%         Evaluation sets summary.
%     }
% \begin{tabular}{|l|l|l|}
% \hline
% \multicolumn{1}{|c|}{Decode set} & \multicolumn{1}{c|}{Corpus} & \multicolumn{1}{c|}{Subset} \\ \hline
% LDC                              & LibriSpeech                 & dev-clean                   \\ \hline
% LDO                              & LibriSpeech                 & dev-other                   \\ \hline
% TED-dev                          & TED-LIUM r2                 & validation                  \\ \hline
% TED-tst                          & TED-LIUM r2                 & test                        \\ \hline
% CV-dev                           & CommonVoice V1              & 10\% from corpus            \\ \hline
% CV-test                          & CommonVoice V1              & 10\% from corpus            \\ \hline
% \end{tabular}
% \label{tab:evalsets}
% \end{table*}

\subsection{ASR systems}
\label{subsec:asr-systems}

The main ASR system is based on the pre-trained LibriSpeech model provided by the ESPNet toolkit \cite{watanabe2018interspeech}.
The model implements the transformer architecture \cite{vaswani2017nips},
takes as input 80-dimensional Mel filter banks (extracted with the Kaldi toolkit \cite{povey2011asru})
and outputs a sequence of tokens.
The token vocabulary has dimension 5000 and is obtained by subword segmentation based on a unigram language model \cite{kudo2018acl}.
The model is trained on the 960h of the LibriSpeech dataset,
which is further augmented using the SpecAugment techniques (time warping, frequency masking, time masking) \cite{park2019interspeech}.
% Trained hybrid CTC attention loss.
For decoding we use a language model, which is also implemented as a transformer and is trained on the LibriSpeech transcriptions and other 14,500 public domain books \cite{panayotov2015icassp}.
The vocabulary of the language model consists of the same 5000 tokens as used by the ASR model.

For the ensemble experiments we re-train the ASR system using the same architecture and data, but different random seeds.
We repeat the process four times obtaining four independent models.
Due to computational constraints, these models were trained for a shorter number of epochs than the main system (10 versus 120),
but we observed that the validation loss function curve began to flatten and that the test performance is reasonable
(5.5\%{\small$\pm$0.4} WER on Libri clean vs 2.7\% obtained by the pre-trained model).

\subsection{Evaluation metrics}
\label{subsec:evaluation-metrics}

Ideally, we want the confidence score to be correlated with the correctness of the transcription,
that is, correct words should have large confidence score, while incorrect ones, low score.
% We are generally interested in ranking words based on the confidence score:
Following previous work \cite{hendrycks2016iclr,corbiere2019nips,malinin2020arxiv},
we employ metrics that are generally used for evaluating binary classifiers,
but which have the discrimination threshold varied.
More precisely, we measure the area under precision-recall curve (AUPR) and the area under receiver operating characteristic curve (AUROC).
However, depending on what we want to focus (correctly or erroneously transcribed words) we obtain different variants:
if we are interested in detecting erroneously transcribed words, we will treat the errors as the positive class;
on the other hand if we are interested in the correctly transcribed words, we will treat the latter as the positive class.
Hence, for AUPR we use two variants AUPR$e$ (when errors are treated as positives) and AUPR$s$ (when correct words are treated as positives).
For AUROC the same value is obtained for either choice, so there is no need to make this distinction.

We do not evaluate calibration, since our methodology is not designed to necessarily yield a probability, but a score that is correlated with the label.
The temperature scaling approach does indeed transform the score to a probability (since it learns the scaling coefficients $\alpha$ and $\beta$),
but the same cannot be said about the other approaches (for example, negative entropy).

\section{Results}
\label{sec:results-and-discussion}

This section presents the experimental results.
We start with an evaluation of features and their aggregations (\S\ref{subsec:feats-and-aggs}),
and then report results for the improved variants
involving temperature scaling, dropout (\S\ref{subsec:temp-scale-and-dropout})
and ensembles (\S\ref{subsec:ensembles}).
% We conclude the section with a discussion.

\subsection{Features and aggregation}
\label{subsec:feats-and-aggs}

\begin{table*}
    \center
    \small
    \setlength{\tabcolsep}{3pt}
    \newcommand{\ii}[1]{\footnotesize{\color{gray}#1}}
    \caption{%
        Confidence scoring results for combinations of features and aggregations on the four test splits. % of four popular benchmark datasets: Libri clean, Libri other, TED and CommonVoice.
        For all three metrics reported (AUPR\textit{e}, AUPR\textit{s}, AUROC) larger values are better.
        We indicate the word error rate of the pre-trained ASR system on each of the dataset by the figures on the right of the name.
        % The figures to the right of the dataset name indicate the word error rate of the base system. 
        % We evaluate the confidence estimation performance in terms of three metrics:
        % area under precision recall curve for error detection (AUPR\textit{e}),
        % area under precision recall curve for success detection (AUPR\textit{s}),
        % area under receiver operator curve (AUROC).
    }
    \begin{tabular}{rlcrrrrrrrrrrrr}
        % Results taken from https://docs.google.com/spreadsheets/d/1OZlmXNNF8pbej5QKw98vtuWOZXOPWoluBQknw2D83XM/edit?usp=sharing
        \toprule
        &     &     & \multicolumn{3}{c}{Libri clean / 2.7\%}
                    & \multicolumn{3}{c}{Libri other / 6.0\%}
                    & \multicolumn{3}{c}{TED / 13.3\%}
                    & \multicolumn{3}{c}{CommonVoice / 28.6\%} \\
        \cmidrule(lr){4-6}
        \cmidrule(lr){7-9}
        \cmidrule(lr){10-12}
        \cmidrule(lr){13-15}
        & feat. & agg.
        & AUPR\it{e} & AUPR\it{s} & AUROC
        & AUPR\it{e} & AUPR\it{s} & AUROC
        & AUPR\it{e} & AUPR\it{s} & AUROC
        & AUPR\it{e} & AUPR\it{s} & AUROC %
        \\
        \midrule
        \ii{1}   & log-proba   & sum & 21.55     & \bf 99.21 & 82.41     & \bf 29.99 & \bf 98.10 & \bf 81.75 & \bf 39.97 & 95.88      & 79.95      & \bf 48.98 & \bf  77.71 & \bf 64.84 \\
        \ii{2}   & log-proba   & min & \bf 21.85 & 99.19     & \bf 82.47 & 28.64     & 98.06     & 81.66     & 39.74     & \bf  95.94 & \bf  80.58 & 46.79     & 76.74      & 62.67 \\
        \ii{3}   & log-proba   & avg & 20.12     & 99.10     & 80.90     & 26.72     & 97.93     & 80.47     & 38.74     & 95.88      & 80.29      & 44.51     & 75.82      & 60.87 \\
        \midrule
        \ii{4}   & neg-entropy & sum & 17.31     & \bf 99.10 & 79.97     & 26.37     & \bf 97.86 & 79.58     & 34.96     & 95.41      & 77.57      & \bf 47.71 & \bf 77.10  & \bf  63.74 \\
        \ii{5}   & neg-entropy & min & \bf 19.94 & 99.09     & \bf 80.55 & \bf 26.75 & 97.82     & \bf 79.64 & \bf 37.55 & \bf  95.56 & \bf 79.01  & 45.51     & 76.00      & 61.21 \\
        \ii{6}   & neg-entropy & avg & 17.55     & 98.95     & 77.72     & 24.26     & 97.59     & 77.46     & 36.28     & 95.42      & 78.29      & 42.64     & 74.83      & 58.75 \\
        % log-proba-pred & sum & n & 23.81 & 99.25 & 83.37 & 28.16 & 98.14 & 82.15 & 41.23 & 96.02 & 80.77 & 48.51 & 77.76 & 64.91 \\
        % log-proba-pred & min & n & 23.66 & 99.22 & 83.34 & 27.16 & 98.11 & 82.05 & 41.15 & 96.06 & 81.25 & 46.19 & 76.79 & 62.77 \\
        % log-proba-pred & avg & n & 22.17 & 99.14 & 81.82 & 25.90 & 98.00 & 81.07 & 40.07 & 96.02 & 81.02 & 44.12 & 75.90 & 61.08 \\
        % \midrule
        % log-proba      & sum & y & 22.38 & 99.26 & 83.45 & 30.35 & 98.20 & 82.54 & 40.92 & 96.19 & 81.11 & 48.67 & 77.39 & 64.33 \\
        % log-proba      & min & y & 21.78 & 99.18 & 82.29 & 28.53 & 98.03 & 81.41 & 39.84 & 95.98 & 80.74 & 46.86 & 76.80 & 62.79 \\
        % log-proba      & avg & y & 19.49 & 99.04 & 79.73 & 26.08 & 97.78 & 79.23 & 38.97 & 95.99 & 80.66 & 44.77 & 76.05 & 61.33 \\
        % log-proba-pred & sum & y & 22.63 & 99.17 & 81.83 & 27.99 & 97.98 & 80.85 & 41.03 & 95.90 & 80.34 & 48.15 & 77.01 & 63.69 \\
        % log-proba-pred & min & y & 22.80 & 99.09 & 80.98 & 26.61 & 97.80 & 79.70 & 40.72 & 95.78 & 80.29 & 45.69 & 76.10 & 61.50 \\
        % log-proba-pred & avg & y & 20.76 & 98.97 & 78.66 & 24.85 & 97.63 & 78.13 & 39.30 & 95.67 & 79.73 & 43.39 & 75.23 & 59.77 \\
        % neg-entropy    & sum & y & 23.60 & 99.38 & 85.54 & 31.01 & 98.53 & 84.80 & 42.16 & 96.91 & 83.50 & 49.17 & 77.88 & 65.09 \\
        % neg-entropy    & min & y & 22.99 & 99.33 & 84.98 & 29.83 & 98.45 & 84.31 & 41.23 & 96.87 & 83.50 & 48.00 & 77.78 & 64.46 \\
        % neg-entropy    & avg & y & 21.49 & 99.21 & 82.90 & 27.91 & 98.20 & 82.50 & 40.22 & 96.53 & 82.48 & 46.04 & 77.00 & 63.04 \\
        \bottomrule
    \end{tabular}
    \label{tab:feat-agg}
\end{table*}

\begin{figure}
    \center
    \includegraphics[width=0.4\textwidth]{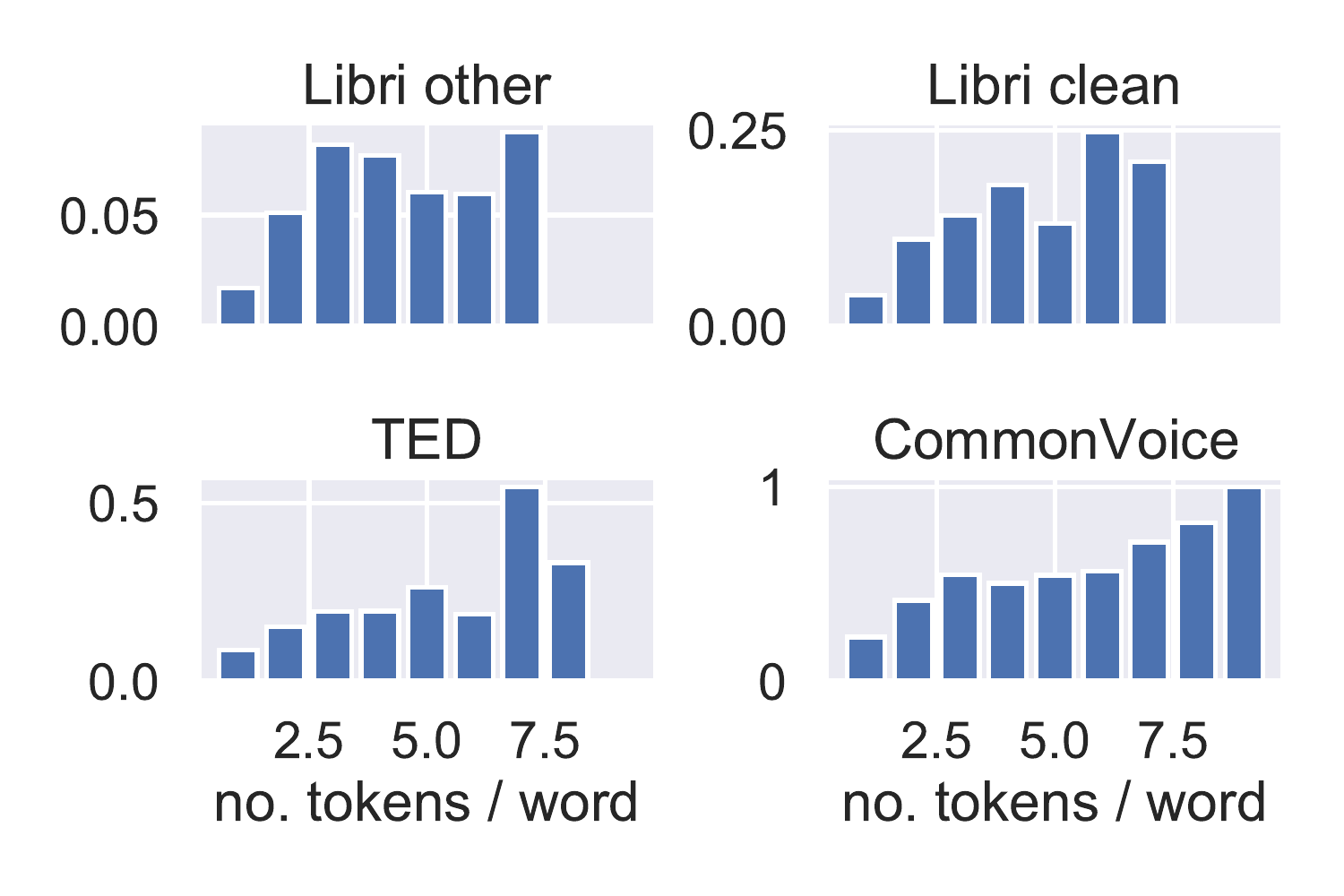}
    \caption{%
        Fraction of errors as a function of the word length.
        The fraction of errors is computed as the number of erroneously transcribed words divided by the total number of words,
        while the word length is measured as number of tokens.
    }
    \label{fig:error-vs-length}
\end{figure}

We evaluate the proposed uncertainty features and aggregation techniques on the four datasets described in subsection \ref{subsec:datasets}.
We use the pre-trained model to obtain text predictions for all the audio files in the test split of each dataset,
and then estimate the confidence based on the methodology described in subsection \ref{subsec:confidence-estimation}.
Table \ref{tab:feat-agg} presents the results for all combinations of features and aggregations.

\textit{Comparison of features.}
We observe that log probability features outperform the entropy features across all settings (aggregations and datasets).
The only notable exception is the CommonVoice dataset where the results are comparable.

\textit{Comparison of aggregations.}
Generally, the sum aggregation works better with log-proba features,
while the min aggregation works better for entropy features.
The sum might not be well suited for entropy features because their magnitude is larger than for log-proba
and the word confidence gets penalized too much by the length;
but, as we will see further, this behaviour can be alleviated by temperature scaling.
Averaging is generally underperforming for both features,
suggesting that length-invariant measures are detrimental.
Indeed, a closer look at the frequency of errors with the length size indicates that the
more tokens a words has the more likely is that it is incorrect, see Figure \ref{fig:error-vs-length}.
Statistical tests (paired $t$-tests on the twelve results from each configuration at $p=0.05$)
confirm that for both features the sum and min aggregation are significantly better than avg,
while the statistical test between sum and min did not reject the null hypothesis for neither feature.

\textit{Comparison across datasets.}
As expected, the pre-trained model performs best on in-domain data (2.7\% WER on Libri clean and 6.0\% on Libri other),
the performance then dropping sharply as we evaluate on out-of-domain data (13.3\% on TED and 28.6\% on CommonVoice).
In each of these settings the number of words that are correctly classified changes,
going from more on the Libri splits to fewer on TED and CommonVoice.
This observation explains why the performance for AUPR$s$ drops as a function of the domain of the data,
and, conversely, why the AUPR$e$ performance improves.
Unfortunately, for this exact reason---the different performance of the base ASR system on the four datasets---%
it is impossible to compare the confidence methods across datasets, as they use a different groundtruth \cite{ashukha2020iclr}.

\subsection{Temperature scaling and dropout}
\label{subsec:temp-scale-and-dropout}

\begin{table}
  \centering
  \small
  \setlength{\tabcolsep}{5pt}
  \caption{%
    Confidence scoring results on the TED test set for combinations of features, aggregations and their improved variants -- %
    temperature scaling (\textsf{TS}) and dropout (\textsf{D}).
    The bullet sign $\bullet$ indicates whether a variant is employed.
    Bold results indicate the best results for the feature-aggregation combination;
    these results show that using both temperature scaling and dropout yields the best results.
  }
  \newcommand{\ii}[1]{\footnotesize{\color{gray}#1}}
  \begin{tabular}{rlcccrrr}
    \toprule
            & feat.                         & agg.                 & \sf TS    & \sf D     & AUPR\it{e} & AUPR\it{s} & AUROC \\
    \midrule
     \ii{1}  & \multirow{4}{*}{log-proba}   & \multirow{4}{*}{sum} &           &           & 39.97     & 95.88     & 79.95 \\
     \ii{2}  &                              &                      &           & $\bullet$ & 41.41     & 96.81     & 82.78 \\
     \ii{3}  &                              &                      & $\bullet$ &           & 40.92     & 96.19     & 81.11 \\
     \ii{4}  &                              &                      & $\bullet$ & $\bullet$ & \bf 42.99 & \bf 97.14 & \bf 84.10 \\
    \midrule
     \ii{5}  & \multirow{4}{*}{log-proba}   & \multirow{4}{*}{min} &           &           & 39.74     & 95.94     & 80.58 \\
     \ii{6}  &                              &                      &           & $\bullet$ & 42.08     & 96.94     & 83.76 \\
     \ii{7}  &                              &                      & $\bullet$ &           & 39.84     & 95.98     & 80.74 \\
     \ii{8}  &                              &                      & $\bullet$ & $\bullet$ & \bf 42.17 & \bf 97.00 & \bf 83.93 \\
    \midrule
     \ii{9}  & \multirow{4}{*}{log-proba}   & \multirow{4}{*}{avg} &           &           & 38.74     & 95.88     & 80.29 \\
    \ii{10}  &                              &                      &           & $\bullet$ & 41.19     & 96.95     & 83.73 \\
    \ii{11}  &                              &                      & $\bullet$ &           & 38.97     & 95.99     & 80.66 \\
    \ii{12}  &                              &                      & $\bullet$ & $\bullet$ & \bf 41.32 & \bf 97.06 & \bf 84.08 \\
    \midrule
    \ii{13}  & \multirow{4}{*}{neg-entropy} & \multirow{4}{*}{sum} &           &           & 34.96     & 95.41     & 77.57 \\
    \ii{14}  &                              &                      &           & $\bullet$ & 33.14     & 96.22     & 79.45 \\
    \ii{15}  &                              &                      & $\bullet$ &           & 42.16     & 96.91     & 83.50 \\
    \ii{16}  &                              &                      & $\bullet$ & $\bullet$ & \bf 43.59 & \bf 97.62 & \bf 85.51 \\
    \midrule
    \ii{17}  & \multirow{4}{*}{neg-entropy} & \multirow{4}{*}{min} &           &           & 37.55     & 95.56     & 79.01 \\
    \ii{18}  &                              &                      &           & $\bullet$ & 38.75     & 96.53     & 81.98 \\
    \ii{19}  &                              &                      & $\bullet$ &           & 41.23     & 96.87     & 83.50 \\
    \ii{20}  &                              &                      & $\bullet$ & $\bullet$ & \bf 42.23 & \bf 97.60 & \bf 85.51 \\
    \midrule
    \ii{21}  & \multirow{4}{*}{neg-entropy} & \multirow{4}{*}{avg} &           &           & 36.28     & 95.42     & 78.29 \\
    \ii{22}  &                              &                      &           & $\bullet$ & 38.01     & 96.51     & 81.85 \\
    \ii{23}  &                              &                      & $\bullet$ &           & 40.22     & 96.53     & 82.48 \\
    \ii{24}  &                              &                      & $\bullet$ & $\bullet$ & \bf 41.15 & \bf 97.43 & \bf 85.18 \\
    \bottomrule
  \end{tabular}
  \label{tab:improve-proba-ts-dropout}
\end{table}

We benchmark the confidence scoring method after improving the token probabilities by two of the described techniques:
temperature scaling and dropout.
We use the pre-trained ASR system and report results only on the TED test set.
The parameters for temperature scaling method are learnt on the \texttt{dev} split of the TED dataset for each setting of feature and aggregation.
When temperature scaling is combined with dropout we first apply the temperature scaling (using the same temperature) and the follow with the aggregation over dropout.
The dropout method averages 64 independent predictions.
Table \ref{tab:improve-proba-ts-dropout} presents the results for all combinations of features and aggregations and improvement techniques.

\begin{figure}
    \centering
    \includegraphics[width=0.35\textwidth]{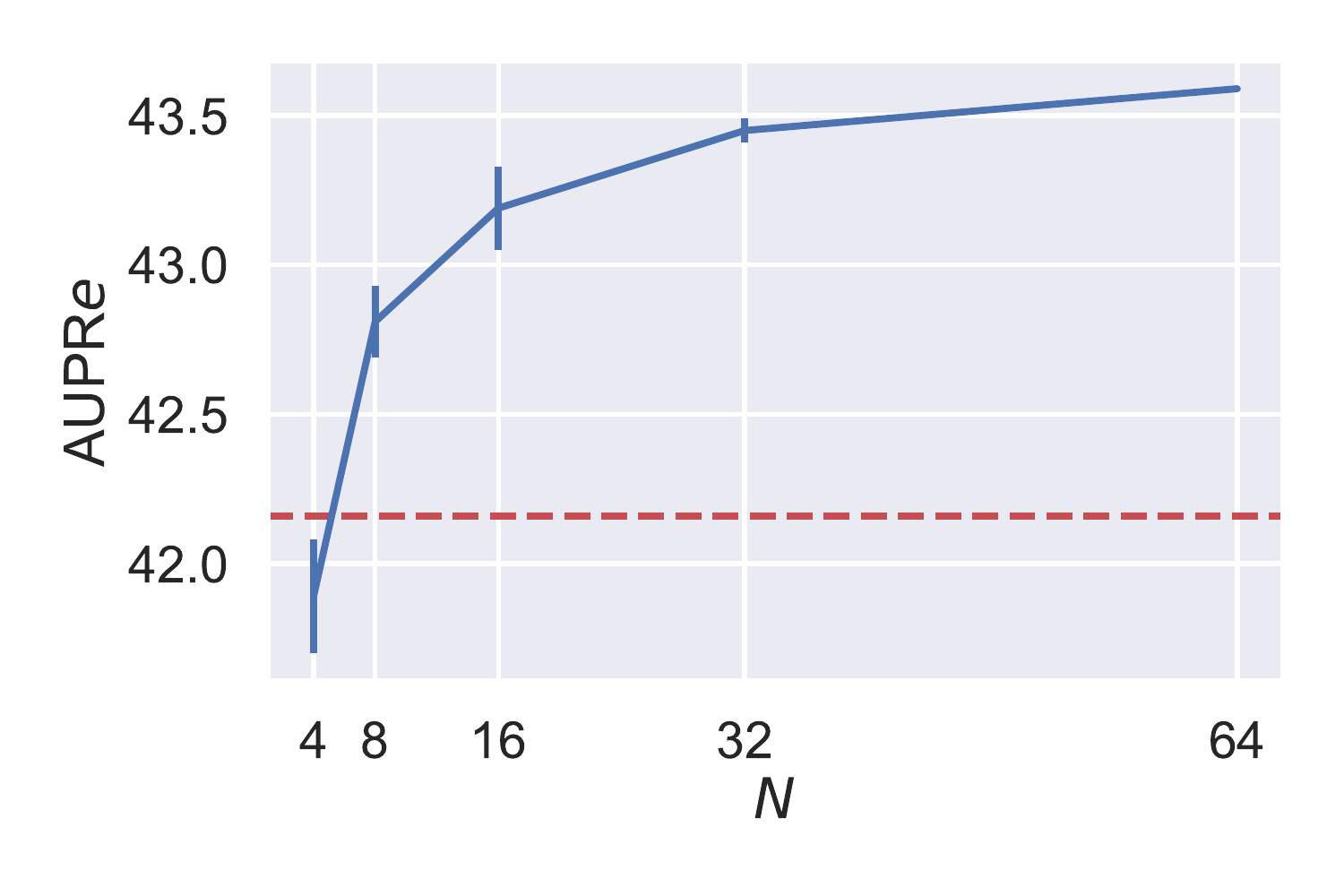}
    \caption{%
        AUPR$e$ performance as a function of the number of dropout runs on the TED test set.
        The horizontal red line indicates the performance of the model without dropout.
        The model uses neg-entropy features, sum aggregation and temperature scaling.
    }
    \label{fig:dropout-n}
\end{figure}

The results indicate that both proposed methods improve the results as is their combination, which gives overall the best result.
We observe that log-proba features benefit more from dropout,
while the neg-entropy feature yield more improvements when temperature scaling is used.
Interestingly, the best results are now obtained for the neg-entropy with sum aggregation (row 16).
Figure \ref{fig:dropout-n} shows that the dropout performance improves with the number of runs
and plateaus around the chosen value of 64.

% TODO calibration plot?

\subsection{Ensembles}
\label{subsec:ensembles}

\begin{table}
  \centering
  \small
  \caption{%
    Confidence scoring results on the TED test set for combinations of
    temperature scaling (\textsf{TS}), dropout (\textsf{D}) and ensembles (\textsf{E}),
    using neg-entropy features and sum aggregation.
    % using neg-entropy as features and sum as aggregation.
    % The bullet sign $\bullet$ indicates whether a variant is employed.
  }
  \newcommand{\ii}[1]{\small{\color{gray}#1}}
  \begin{tabular}{rccccrrr}
    % bash evaluate_ensembles.sh ted-medium
    \toprule
           & \sf TS      & \sf D     & \sf E & & AUPR\it{e} & AUPR\it{s} & AUROC \\
    \midrule
    \ii{1}   &           &           &           &  & 28.58     & 95.30     & 75.79 \\ % 26.55     & 95.73 & 75.61\\
    \midrule
    \ii{2}   & $\bullet$ &           &           &  & 32.00     & 96.32     & 79.47 \\ % 30.07     & 96.70 & 79.52 \\
    \ii{3}   &           & $\bullet$ &           &  & 27.49     & 95.51     & 75.67 \\ % 25.18     & 95.82 & 75.28 \\
    \ii{4}   &           &           & $\bullet$ &  & 30.89     & 96.26     & 78.89 \\ % 28.86     & 96.44 & 78.62\\
    \midrule
    \ii{5}   & $\bullet$ & $\bullet$ &           &  & 31.10     & 96.40     & 79.06 \\ % 29.17     & 96.68 & 78.98 \\
    \ii{6}   & $\bullet$ &           & $\bullet$ &  & \bf 34.57 & \bf 96.95 & \bf 81.64 \\ % 32.92 & 97.15 & 81.58 \\
    \ii{7}   &           & $\bullet$ & $\bullet$ &  & 28.94     & 96.26     & 77.93 \\ % \\
    \midrule
    \ii{8}   & $\bullet$ & $\bullet$ & $\bullet$ &  & 33.00     & 96.84     & 80.82 \\ % \\
    \bottomrule
  \end{tabular}
  \label{tab:improve-proba-ts-dropout-ensemble}
\end{table}

We present results for confidence scoring using ensembles of models and their combinations with the other improved versions (temperature scaling and dropout).
For each of the retrained models from the ensemble we use the predictions of the pre-trained model to select the transcription;
the retrained model is just used for confidence scoring, by extracting the confidence features described previously.
The results are presented in Table \ref{tab:improve-proba-ts-dropout-ensemble}.
For the rows that do not use ensemble (rows 1, 2, 3 and 5) we evaluate each of the four single models independently and report the mean performance.

The pre-trained model (Table \ref{tab:improve-proba-ts-dropout}, row 13) has generally a better performance than the retrained ones (Table \ref{tab:improve-proba-ts-dropout-ensemble}, row 1),
suggesting that the predictive performance of a model can correlate with its confidence scoring performance.

Among the three improvement methods, we note that temperature scaling gives the largest performance boost on all three metrics (row 2).
Surprisingly, the dropout method improves only the AUPR$s$ performance over the baseline (row 3).
On combinations of two methods, temperature scaling and ensemble complement each other and obtain better performance.

% \subsection{Discussion}
% \label{subsec:discussion}
% 
% We briefly discuss different perspectives on our work.
% 
% \textit{Augmenting the feature set}.
% The benchmarked methods have the benefit of being general,
% as we leverage the posterior token predictions,
% which are readily available in most, if not all, existing end-to-end ASR toolkits.
% However, the feature set could be extended with prior probabilities on the input audio or the generated text,
% or with duration information extracted from the attention weights.
% 
% \textit{Learning features.}
% Following \cite{corbiere2019nips, chen2019aistats},
% we have also experimented with learning a confidence scoring network on top of features extracted from the end-to-end model (specifically, logits and pre-logits activations).
% However, our experiments failed to show improvements over the presented results.
% 
% \textit{Dealing with deletions.}
% To generate confidence scoring groundtruth we align the reference text to the predicted text
% and mark the correct words in the predicted text as positives and the substitutions and insertions as negatives.
% This approach is typical in the confidence scoring literature, but it misses the errors made by deleting words.
% Several works have addressed this problem \cite{seigel2014icassp,ragni2018slt},
% and we leave it to future work to extend our approach for this task.

\section{Conclusions}
\label{subsec:conclusion}

This paper presented an approach for word-level confidence scoring in end-to-end speech recognition systems.
We carried a thorough ablation study on features and their aggregation on three well-known speech databases (LibriSpeech, TED-LIUM and CommonVoice)
and further evaluated improved methods, which modify the token probabilities, and their combinations.
Our main observation is that temperature scaling improves both uncertainty features (log-proba and neg-entropy)
as well as the other two methods (dropout and ensemble).
Using a pre-trained model allows replicability and enables comparison with future confidence scoring methods that will use the same ASR.
% as they can be evaluated in the same setting (by keeping the ASR model fixed).
We strived for simplicity by using a compact feature set (based on readily-available token posteriors);
in future work we will consider augmenting these features with complementary information (\eg, token duration extraction from attention).
% Another 

\section{Acknowledgements}
This work was supported by
% TADARAV
the PCCDI UEFISCDI project (funded by the Romanian Ministry of Research and Innovation, PN-III-P1-1.2-PCCDI-2017-0818/73) and
%{\footnotesize This work was partially supported by a grant of the Romanian Ministery of Research and Innovation, CCCDI – UEFISCDI, project number PN-III-P1-1.2-PCCDI-2017-0818 / 73, within PNCDI III.}
% Alexandru`s Caranica ACK:
the POCU project (funded by the Romanian Ministry of European Funds, financial agreement 51675/09.07.2019, SMIS code 125125).
%{\footnotesize The work was also partially funded by the Operational Programme Human Capital of the Ministry of European Funds through the Financial Agreement 51675/09.07.2019, SMIS code125125.}

\bibliographystyle{IEEEbib}
\bibliography{myref}

\end{document}